\documentclass[a4paper]{spie}  

\usepackage[]{graphicx}

\title{Spin-Fluctuation Mechanism of Superconductivity in Cuprates} 

\author{A. Ram\v sak\supit{a,b} and P. Prelov\v sek\supit{a,b} 
\skiplinehalf
\supit{a}Faculty of Mathematics and Physics,
  University of Ljubljana, Ljubljana, Slovenia; \\
\supit{b}J.\ Stefan Institute, Ljubljana, Slovenia
}

\authorinfo{Further information: anton.ramsak@ijs.si}

\begin{document} 
  \maketitle 

\begin{abstract}
Normal and superconducting state spectral properties of cuprates are
theoretically described within the extended $t$-$J$ model. The method
is based on the equations of motion for projected fermionic operators
and the mode-coupling approximation for the self-energy matrix. The
dynamical spin susceptibility at various doping is considered as an
input, extracted from experiments. The analysis shows that the onset of
superconductivity is dominated by the spin-fluctuation
contribution. The coupling to spin fluctuations directly involves the
next-nearest-neighbor hopping $t'$, hence $T_c$ shows a pronounced
dependence on $t'$. The latter can offer an explanation for the
variation of $T_c$ among different families of hole-doped cuprates. A
formula for maximum $T_c$ is given and it is shown that optimum
doping, where maximum $T_c$ is reached, is with increasing $-t'$ 
progresively increased.
\end{abstract}



\section{INTRODUCTION}
\label{sect:intro}  

The mechanism of high-temperature supeconductivity (SC) in cuprates
represents one of the central open questions in the solid state
theory.  The role of strong correlations and the antiferromagnetic
(AFM) state of the reference insulating undoped compound has been
recognized very early \cite{ande}. Still, up to date there is no
general consensus whether ingredients as embodied within the prototype
single-band models of strongly correlated electrons are sufficient to
explain the onset of high $T_c$, or in addition other degrees of
freedom, as e.g. phonons, should be invoked. As the basis of our
study, we assume the extended $t$-$J$ model \cite{bask}, allowing for
the next-nearest-neighbor (NNN) hopping $t'$ term. The latter model,
as well as the Hubbard model \cite{bulu}, both closely related in the
strong correlation limit $U \gg t$, have been considered by numerous
authors to address the existence of SC due to strong correlations
alone.  Within the parent resonating-valence-bond (RVB) theory
\cite{ande,bask,gros} and slave-boson approaches to the $t$-$J$ model
\cite{kotl} the SC emerges due to the condensation of singlet pairs,
induced by the exchange interaction $J$.  An alternative view on
strong correlations has been that AFM spin fluctuations, becoming
particularly longer-ranged and soft at low hole doping, represent the
relevant low-energy bosonic excitations mediating the attractive
interaction between quasiparticles (QP) and induce the $d$-wave SC
pairing. The latter scenario has been mainly followed in the planar
Hubbard model \cite{bulu} and in the phenomenological spin-fermion
model \cite{mont}.  Recent numerical studies of the planar $t$-$J$
model using the variational quantum Monte Carlo approach \cite{sore},
as well as of the Hubbard model using cluster dynamical mean-field
approximation \cite{maie}, seem to confirm the stability of the
$d$-wave SC as the ground state at intermediate hole doping.  The
relevance of $t'$ for $T_c$ has been already recognized \cite{fein}
and recently, there are also some numerical studies of the influence
of $t'$ on pairing within prototype models \cite{whit,shih}, although
with conflicting conclusions.

\section{MODEL AND METHOD} 

The $t$-$J$ model is nonperturbative by construction, so it is hard to
design for it a trustworthy analytical method. One approach is to use
the method of equations of motion (EQM) to derive an effective
coupling between fermionic QP and spin fluctuations \cite{prel}. The
latter method has been employed to evaluate the self energy and
anomalous properties of the spectral function \cite{prel,plak,prel1},
in particular the appearance of the pseudogap and the effective
truncation of the Fermi surface (FS) at low hole doping
\cite{prel1}. The analysis has been extended to the study of the SC
pairing \cite{plak,prel2}, while an analogous approach has been also applied
to the Hubbard model \cite{onod}.

\subsection{Equations of motion}

We employ the formalism of the EQM and the resulting
Eliashberg equations within the simplest mode-coupling approximation
\cite{plak,prel1,prel2}. Equations involve the dynamical spin susceptibility
which we consider as an phenomenological input taken from the
inelastic-neutron-scattering (INS) and NMR-relaxation experiments in
cuprates.  The analysis of these experiments \cite{bonc} reveals that
in the metallic state the AFM staggered susceptibility is strongly
enhanced at the crossover from the overdoped (OD) regime to optimum
(OP) doping and is increasing further in underdoped (UD) cuprates,
while at the same time the corresponding spin-fluctuation energy scale
is becoming very soft. Direct evidence for the latter is the
appearance of the resonant magnetic mode \cite{ross,fong} within the
SC phase indicating that the AFM paramagnon mode can become even lower
than the SC gap.  These facts give a support to the scenario that spin
fluctuations in cuprates represent the lowest bosonic mode relevant
for the $d$-wave SC pairing. 

One of the central results of our EQM approach is that the relevant
coupling to AFM paramagnons involves directly $t'$, but not $t$. The
evident consequence is the sensitivity of $T_c$ on $t^\prime$,
consistent with the experimental evidence for different families of
cuprates \cite{pava}.

We consider the extended $t$-$J$ model
\begin{equation}
H=-\sum_{i,j,s}t_{ij} \tilde{c}^\dagger_{js}\tilde{c}_{is}
+J\sum_{\langle ij\rangle}({\bf S}_i\cdot {\bf S}_j-\frac{1}{4}
n_in_j), \label{tj} 
\end{equation}
including both the NN hopping $t_{ij}=t$ and the NNN hopping
$t_{ij}=t^\prime$. The projection in fermionic operators,
$\tilde{c}_{is}= (1-n_{i,-s}) c_{is}$ leads to a nontrivial EQM, which
can be in the ${\bf k}$ basis represented as
\begin{eqnarray}
\label{eqm}
&&[\tilde c_{{\bf k} s},H]= [(1+c_h) \frac{\epsilon^0_{\bf k}}{2} - J
(1-c_h)] \tilde c_{{\bf k} s} + \\ &&\frac{1}{\sqrt{N}}\sum_{\bf q}
m_{\bf k q} \bigl[ s S^z_{\bf q} \tilde c_{{\bf k}-{\bf q},s} +
S^{\mp}_{\bf q} \tilde c_{{\bf k}-{\bf q},-s} -
\frac{1}{2} \tilde n_{\bf q} \tilde c_{{\bf k}-{\bf q}, s}\bigr],
\nonumber
\end{eqnarray}
where $c_h$ is the hole concentration and $m_{\bf k q}$ is the
effective spin-fermion coupling $ m_{\bf k q}=2J \gamma_{\bf q} +
\epsilon^0_{{\bf k}-{\bf q}}$, while $\epsilon^0_{\bf
  k}=-4t\gamma_{\bf k}- 4t'\gamma'_{\bf k}$ is the bare band
dispersion on a square lattice.  We use the symmetrized coupling as
derived in Ref.~\citenum{prel1} to keep a similarity with the
spin-fermion phenomenology \cite{mont}
\begin{equation}
\label{mkq}
\tilde m_{\bf kq}= 2J \gamma_{\bf q}+
\frac{1}{2} (\epsilon^0_{{\bf k}-{\bf q}}+\epsilon^0_{\bf k}).
\end{equation}
EQM, Eq.~(\ref{eqm}), are used to derive the approximation for the
Green's function (GF) matrix $G_{{\bf k}s}(\omega)= \langle\langle
\Psi_{{\bf k}s}| \Psi^\dagger_{{\bf k}s} \rangle\rangle_\omega$ for
the spinor $\Psi_{{\bf k}s}=(\tilde c_{{\bf k},s},\tilde
c^\dagger_{-{\bf k},-s})$.  

\subsection{Gap equation}

We follow the method, as applied to the
normal state (NS) GF by present authors \cite{prel,prel1}, and
generalized to the SC pairing in Ref.~\citenum{plak}. In
general, we can represent the GF matrix in the form
\begin{equation}
G_{{\bf k}s}(\omega)^{-1}=\frac{1}{\alpha} [\omega \tau_0 -\hat
\zeta_{{\bf k}s} +\mu \tau_3 -\Sigma_{{\bf k}s}(\omega) ], \label{gf}
\end{equation}
where $\alpha = \sum_i \langle \{\tilde c_{i s},\tilde c^{\dagger}_{i
s}\}_+ \rangle/N = (1+c_h)/2$ is the normalization factor, $\mu$ is
the chemical potential and the frequency matrix, 
$\hat \zeta_{{\bf k}s}=\frac{1}{\alpha}\langle \{[\Psi_{{\bf k}s},H],
\Psi^{\dagger}_{{\bf k}s} \}_+ \rangle$,
which generates a renormalized band $\tilde \zeta_{\bf k}=
\zeta^{11}_{{\bf k}s}= \bar \zeta - 4 \eta_1 t \gamma_{\bf k}-
4 \eta_2 t' \gamma'_{\bf k}$ and the mean-field (MF) SC gap
\begin{equation}
\Delta^0_{\bf k}=\zeta^{12}_{{\bf k}s}= - \frac{4J}{N\alpha} 
\sum_{\bf q} \gamma_{{\bf k} -{\bf q}} \langle 
\tilde c_{-{\bf q},-s} \tilde c_{{\bf q},s} \rangle.
\label{del0}
\end{equation}

To evaluate $\Sigma_{{\bf k}s}(\omega)$ we use the lowest-order
mode-coupling approximation, analogous to the treatment of the SC in
the spin-fermion model \cite{mont}, introduced in the $t$-$J$ model
for the NS GF \cite{prel,prel1} and extended to the analysis of the SC
state \cite{plak}. Taking into account EQM, Eq.~(\ref{eqm}), and by
decoupling fermionic and bosonic degrees of freedom, one gets
\begin{equation}
\Sigma^{11(12)}_{{\bf k}s}(i \omega_n)=\frac{-3}{N\alpha\beta}
\sum_{{\bf q},m} \tilde m^2_{\bf kq} G^{11(12)}_{{\bf k}-{\bf
q},s}(i \omega_m) \chi_{\bf q}(i \omega_n-i \omega_m) 
\label{sig}
\end{equation}
where $i \omega_n=i \pi(2n+1)/\beta$ and $\chi_{\bf q}(\omega)$ is the
dynamical spin susceptibility, whereby we have neglected the
charge-fluctuation contribution.

In order to analyze the low-energy behavior in the NS and in the SC
state, we use the QP approximation for the spectral function matrix
\begin{equation}
A_{{\bf k}s}(\omega) \sim \frac{\alpha Z_{\bf k}} {2 E_{\bf k}}
(\omega \tau_0 -\epsilon_{\bf k}\tau_3 -
\Delta_{{\bf k}s}\tau_1) [\delta(\omega-E_{\bf k} )-
\delta(\omega+E_{\bf k})], \label{ak}
\end{equation}
where $E_{\bf k}=(\epsilon^2_{\bf k}+\Delta^2_{{\bf k}s})^{1/2}$,
while NS parameters, i.e., the QP weight $Z_{\bf k}$ and the QP
energy $\epsilon_{\bf k}$, are determined from $G_{{\bf k}s}(\omega
\sim 0)$, Eq.~(\ref{gf}).  The renormalized SC gap is
\begin{equation}
\Delta_{{\bf k}s}= Z_{\bf k}[\Delta^0_{\bf k}+\Sigma^{12}_{{\bf k}s}(0)].
\end{equation}
It follows from Eq.~(\ref{gf}) that $G^{12}_{{\bf k}s}(i\omega_n) \sim
- \alpha Z_{\bf k} \Delta_{{\bf k}s}/(\omega_n^2+E^2_{\bf k})$.  By
defining the normalized frequency dependence $F_{\bf
q}(i\omega_l)=\chi_{\bf q}(i\omega_l)/\chi^0_{\bf q}$, and rewriting
the MF gap, Eq.~(\ref{del0}), in terms of the spectral function,
Eq.~(\ref{ak}), we can display the gap equation in a more familiar
form,
\begin{eqnarray}
\Delta_{{\bf k}s}&=&\frac{1}{N}\sum_{\bf q}[4J \gamma_{{\bf k}-{\bf
q}} -3\tilde m^2_{{\bf k},{\bf k}-{\bf q}}\chi^0_{{\bf k}-{\bf q}}
C_{{\bf q},{\bf k}-{\bf q}}] \times \nonumber \\ &&(Z^0_{\bf k}
Z^0_{\bf q} \Delta_{{\bf q}s} /2 E_{\bf q} ) \mathrm{th}(\beta E_{\bf
q}/2), \label{del}
\end{eqnarray}
where $C_{\bf k q}=I_{\bf kq}(i\omega_n \sim 0)/I^0_{\bf k}$ plays the
role of the cutoff function with
\begin{equation}
I_{\bf kq}(i\omega_n)=\frac{1}{\beta} \sum_{m} F_{\bf q}(i\omega_n-
i\omega_m) \frac{1}{\omega_m^2 + E^2_{{\bf k}s}}, \label{ikq}
\end{equation}
and $I^0_{\bf k}=\mathrm{th}(\beta E_{\bf k}/2)/(2 E_{\bf k})$.
Eq.~(\ref{del}) represents the BCS-like expression which we use
furtheron to evaluate $T_c$, as well as to discuss the SC gap
$\Delta_{\bf q}(T=0)$. To proceed we need the input of two kinds: a)
the dynamical spin susceptibility $\chi_{\bf q}(\omega)$, and b) the
NS QP properties $Z_{\bf k},\epsilon_{\bf k}$.

\subsection{Parameters}

The INS experiments show that within the NS the low-$\omega$ spin
dynamics at ${\bf q} \sim {\bf Q}$ is generally overdamped in the
whole doping (but paramagnetic) regime \cite{fong}. Hence we assume
$\chi_{\bf q}(\omega)$ of the form
\begin{equation}
\chi^{\prime\prime}_{\bf q}(\omega)= \frac{B_{\bf q} \omega}
{\omega^2+\Gamma^2_{\bf q}}, \qquad
F_{\bf q}(i\omega_l)= \frac{\Gamma_{\bf q}}{|\omega_l|+
\Gamma_{\bf q} }. \label{chi}
\end{equation}
Following the recent memory-function analysis \cite{sega} $B_{\bf
q}=\chi^0_{\bf q}\Gamma_{\bf q}$ should be quite independent of
$\tilde {\bf q}={\bf q}-{\bf Q}$. We choose the variation as
$\Gamma_{\bf q} \sim \Gamma_{\bf Q}(1+ w \tilde q^2/\kappa^2)^2$
consistent with the INS observation of faster than Lorentzian fall-off
of $\chi^{\prime\prime}_{\bf q}(\omega)$ vs. $\tilde q$
\cite{fong}. $w \sim 0.42$ in order that $\kappa$ represents the usual
inverse AFM correlation length.

Consequently, we end up with parameters $\chi^0_{\bf Q},\Gamma_{\bf
  Q},\kappa$, which are dependent on $c_h$, but in general as well
vary with $T$. Although one can attempt to calculate them using the
analogous framework \cite{sega}, we use here the experimental input
for cuprates. We refer to results of the recent analysis \cite{bonc},
where NMR $T_{2G}$ relaxation and INS data were used to extract
$\kappa$, $\chi^0_{\bf Q}(T)$ and $\Gamma_{\bf Q}(T)$ for various
cuprates, ranging from the UD to the OD regime. For comparison with
the $t$-$J$ model, we use usual parameters $t=400$~meV, $J=0.3 t$. At
least for UD cuprates, quite consistent estimates for $\chi^0_{\bf
  Q},\Gamma_{\bf Q}$ can be obtained also directly from the INS
spectra \cite{fong}. For UD, OP and OD regime, i.e., $c_h=0.12, 0.17,
0.22$, respectively, we use furtheron the following values:
$\chi^0_{\bf Q}t=15.0, 4.0, 1.0$, $\Gamma^0_{\bf Q}/t=0.03,0.1,0.18$
(appropriate at low $T$), and $\kappa=0.5,1.0,1.2$. It is evident,
that in the UD regime the energy scale $\Gamma^0_{\bf Q}$ becomes very
small (and consequently $\chi^0_{\bf Q} \propto 1/\Gamma^0_{\bf Q}$
large, in spite of modest $\kappa$ \cite{bonc}), supported by a
pronounced resonance mode \cite{fong}. We take into account also the
$T$ dependence, i.e., $\Gamma_{\bf Q}(T) \sim \Gamma^0_{\bf Q} + T$
\cite{bonc}, being significant only in the UD regime.

\section{NUMERICAL RESULTS}
\subsection{Normal state}

For the NS $A_{\bf k}(\omega)$ and corresponding $Z_{\bf
  k},\epsilon_{\bf k}$ we solve Eq.~(\ref{sig}) for $\Sigma^{11}_{\bf
  k}=\Sigma_{\bf k}$ as in Ref.~\citenum{prel1}, with the input for
$\chi_{\bf q}(\omega)$ as described above. Since our present aim is on
the mechanism of the SC, we do not perform the full self-consistent
calculation of $\Sigma_{\bf k}(\omega)$, but rather simplify it as
done in the previous study \cite{prel1}. Large incoherent $\Sigma_{\bf
  k}(\omega \ll 0)$ leads to an overall decrease of the QP weight
$\bar Z<1$ and the QP dispersion with renormalized $\eta_1,\eta_2<1$,
which we assume here following Ref.~\citenum{prel1}, as
$\eta_1=\eta_2=0.5,\bar Z=0.7$. Soft AFM fluctuations with ${\bf q}
\sim {\bf Q}$ lead through Eq.~(\ref{sig}) to an additional reduction
of $Z_{\bf k}$, which is ${\bf k}$-dependent. A pseudogap appears
along the AFM zone boundary and the FS is effectively truncated in the
UD regime with $Z_{{\bf k}_F} \ll 1$ near the saddle points $(\pi, 0)$
(in the antinodal part of the FS) \cite{prel1}. We fix $\mu$ with the
FS volume corresponding to band filling $1-c_h$.

The coupling to low-energy AFM fluctuations, Eq.~(\ref{sig}),
leads to an additional QP renormalization. For fixed $t'/t=-0.3$, we
present in Fig.~1 results for the variation of the $Z_{\bf k}$ in the
Brillouin zone for two sets of parameters, representing the UD and the
OD regime, respectively. The location of the renormalized FS is also
presented in Fig.~1. While the coupling to AFM fluctuations partly
changes the shape of the FS, more pronounced effect is on the QP
weight. It is evident from Fig.~1 that $Z_{\bf k}$ is reduced along
the AFM zone boundary away from the nodal points.  Particularly strong
renormalization $Z_{\bf k} \ll 1 $ happens in the UD case, leading to
an effective truncation of the FS away from nodal points \cite{prel1}.

   \begin{figure}
   \begin{center}
   \begin{tabular}{c}
   \includegraphics[height=7cm]{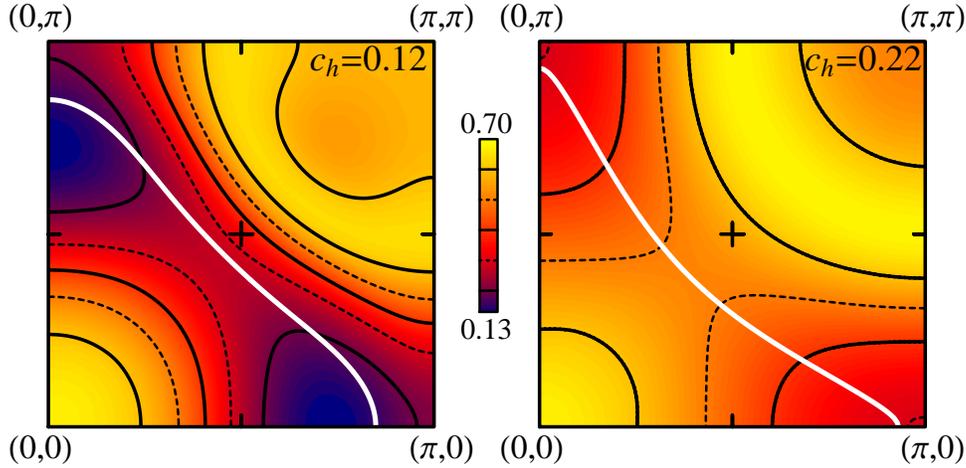}
   \end{tabular}
   \end{center}
   \caption[example] 
   { \label{fig:fig1} 
 QP weight $Z_{\bf k}$ evaluated for $t'/t=-0.3$ 
for parameters corresponding to $c_h=0.12$ and $c_h=0.22$,
respectively. White line represents the location of the FS.
}
   \end{figure} 

\subsection{Superconducting state}

First we comment on the general properties of the gap equation,
Eq.~(\ref{del}). Close to half-filling and for $\chi^0_{\bf q}$ peaked
at ${\bf q}\sim {\bf Q}$ both terms favor the $d_{x^2-y^2}$ SC. The
MF-part $\Delta^0_{\bf k}$, Eq.~(\ref{del0}), involves only $J$ which
induces a nonretarded local attraction, playing the major role in the
RVB theories \cite{ande,bask}. In contrast, the spin-fluctuation part
represents a retarded interaction due to the cutoff function $C_{{\bf
k}{\bf q}}$ determined by $\Gamma_{{\bf k}-{\bf q}}$. The largest
contribution to the SC pairing naturally arises from the antinodal
part of the FS. Meanwhile, in the same region of the FS also $Z_{\bf
k}$ is smallest, reducing the pairing strength in particular in the UD
regime.  Our analysis is also based on the lowest order mode-coupling
treatment of the SC pairing as well as of the QP properties near the
FS. Taking this into account, one can question the relative role of
the hopping parameters $t,t^\prime$ and the exchange $J$ in the
coupling, Eq.~(\ref{mkq}). While our derivation within the $t$-$J$
model is straightforward, an analogous analysis within the Hubbard
model using the projections to the lower and the upper Hubbard band,
respectively, would not yield the $J$ term within the lowest order
since $J \propto t^2$. This stimulates us to investigate in the
following also separately the role of $J$ term in Eq.~(\ref{del}),
both through the MF term, Eq.~(\ref{del0}), and the coupling $\tilde
m_{\bf kq}$, Eq.~(\ref{mkq}).

NS results for $Z_{\bf k},\epsilon_{\bf k}$ are used as an input for
the solution of the gap equation, Eq.~(\ref{del}), as presented in
Fig.~2. For the same $t^\prime/t=-0.3$ we calculate $T_c/t$ for
$c_h=0.12,0.17,0.22$. Besides the result a) of Eq.~(\ref{del}) (full
line in Fig.~2) we present also two alternatives: b) the solution of
Eq.~(\ref{del}) without the MF term, and c) the result with $\tilde
m_{\bf kq}$ without the $J$ term and omitted MF term. In the latter
case, we used as input NS QP parameters, recalculated with
correspondingly modified $\tilde m_{\bf kq}$.

From Fig.~2 it is evident that the spin-fluctuation contribution is
dominant over the MF term. When discussing the role of the $J$ term in
the coupling, Eq.~(\ref{mkq}), we note that in the most relevant
region, i.e., along the AFM zone boundary $\tilde m_{\bf
kQ}=2J-4t^\prime \cos^2 k_x$. Thus, for hole doped cuprates,
$t^\prime<0$ and $J$ terms enhance each other in the coupling, and
neglecting $J$ in $\tilde m_{\bf kq}$ reduces $T_c$, although at the
same time relevant $Z_{\bf k}$ is enhanced.

   \begin{figure}
   \begin{center}
   \begin{tabular}{c}
   \includegraphics[height=7cm]{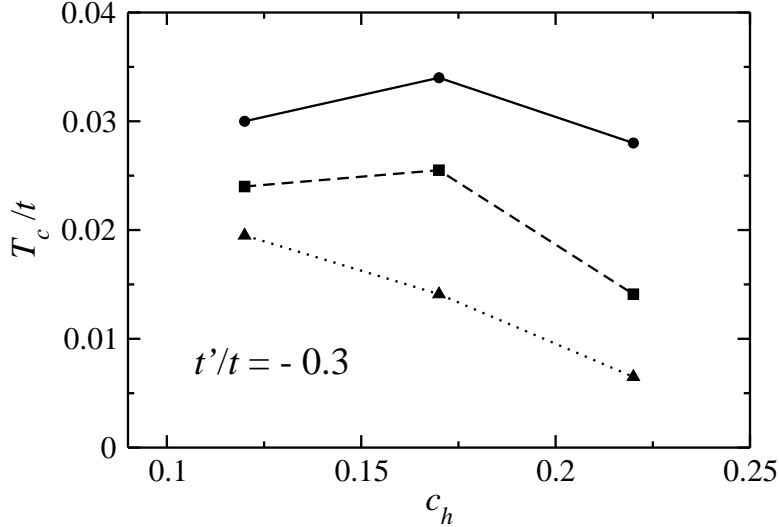}
   \end{tabular}
   \end{center}
   \caption[example] 
   { \label{fig:fig2} 
$T_c/t$ vs. doping $c_h$ for $t'/t=-0.3$, calculated for
various versions of Eq.~(\ref{del}): a) full result (full line), b)
with neglected MF term (dashed line), and c) in addition to b)
modified $\tilde m_{\bf kq}$ without the $J$ term (dotted line).
}
   \end{figure} 

The gap equation Eq.~(\ref{del}) leads to $\Delta_{\bf k}$ with
expected $d_{x^2-y^2}$ symmetry form $\Delta_{\bf k} \sim
\Delta_0(\cos k_x-\cos k_y)/2$, with $\Delta_0(T=0)\sim \eta T_c$ and
$\eta \sim 2.5$. However, we observe that in the UD regime the
effective coherence length $\xi \sim v_F/\Delta_0(T=0)$ becomes very
short. I.e., with $v_F$ taken as the average velocity over the region
$\kappa$ at the antinodal part of the FS we get $\xi$ ranging from
$\xi = 4.4$ in the OD case, to $\xi=1.3$ in the UD example. In the
latter case, SC pairs are quite local and the BCS-like approximation
without phase fluctuations, Eqs.~(\ref{del}),(\ref{tc}), overestimates
$T_c$. Starting from this side, a more local approach would be
desirable. Full numerical solution of Eq.~(\ref{del}) is due to strong
momentum dependent coupling also momentum dependent beyond the simple
$d$-wave form. In Fig.~3(a) $\Delta_{{\bf k}s}(T=0)$ is presented
throughout the Brillouin zone for underdoped $c_h=0.12$ and coupling
with emitted MF term as in (b) from Fig.~2 and for $t'/t=-0.3$. In the
overdoped regime, $c_h=0.22$, the momentum dependence of $\Delta_{{\bf
k}s}$ is less pronounced, Fig.~3(b), in accord with moderate momentum
dependence of $Z_{\bf k}$, Fig.~1(b).

   \begin{figure}
   \begin{center}
   \begin{tabular}{c}
   \includegraphics[height=8cm,angle=-90]{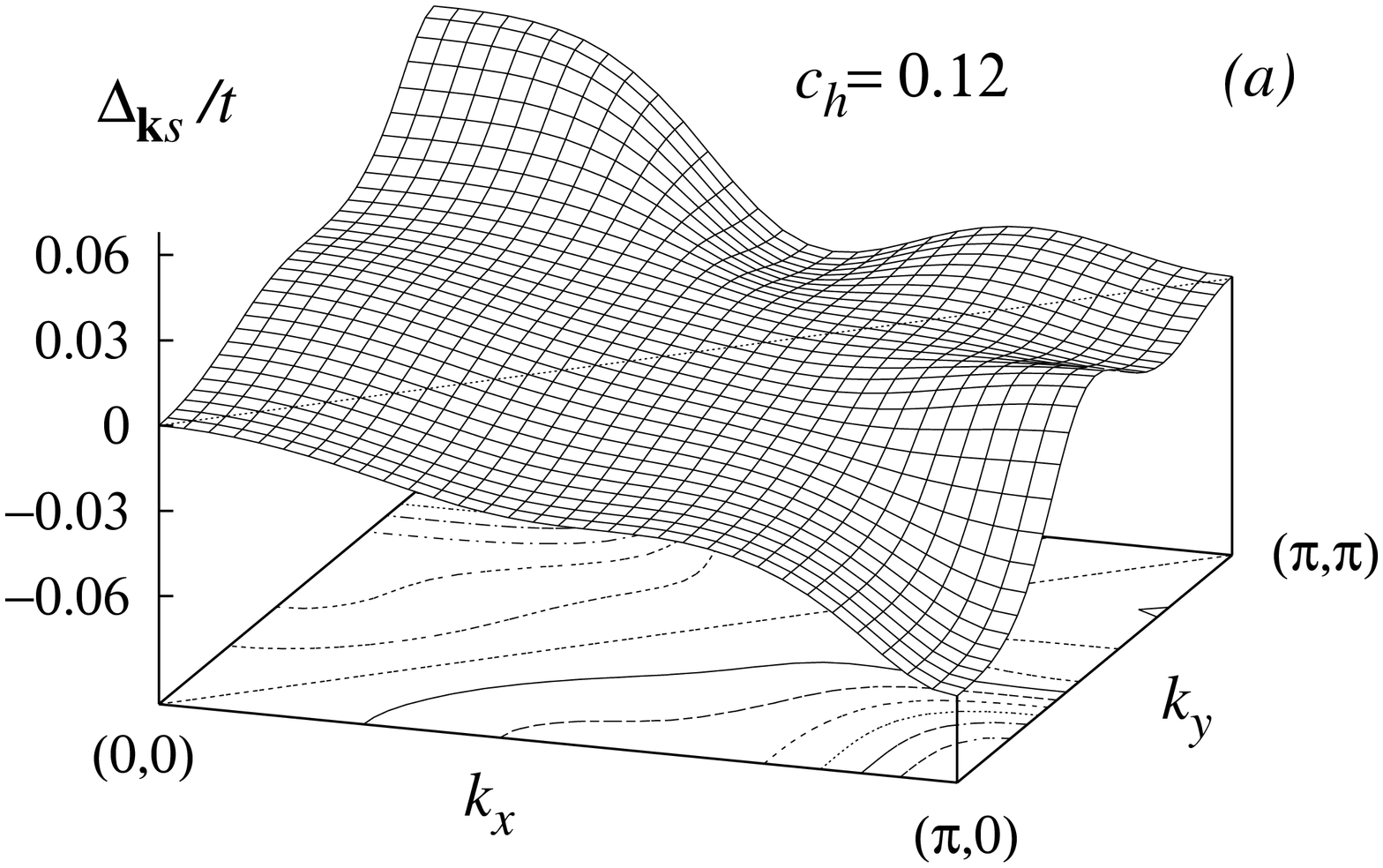}
   \end{tabular}
   \begin{tabular}{c}
   \includegraphics[height=8cm,angle=-90]{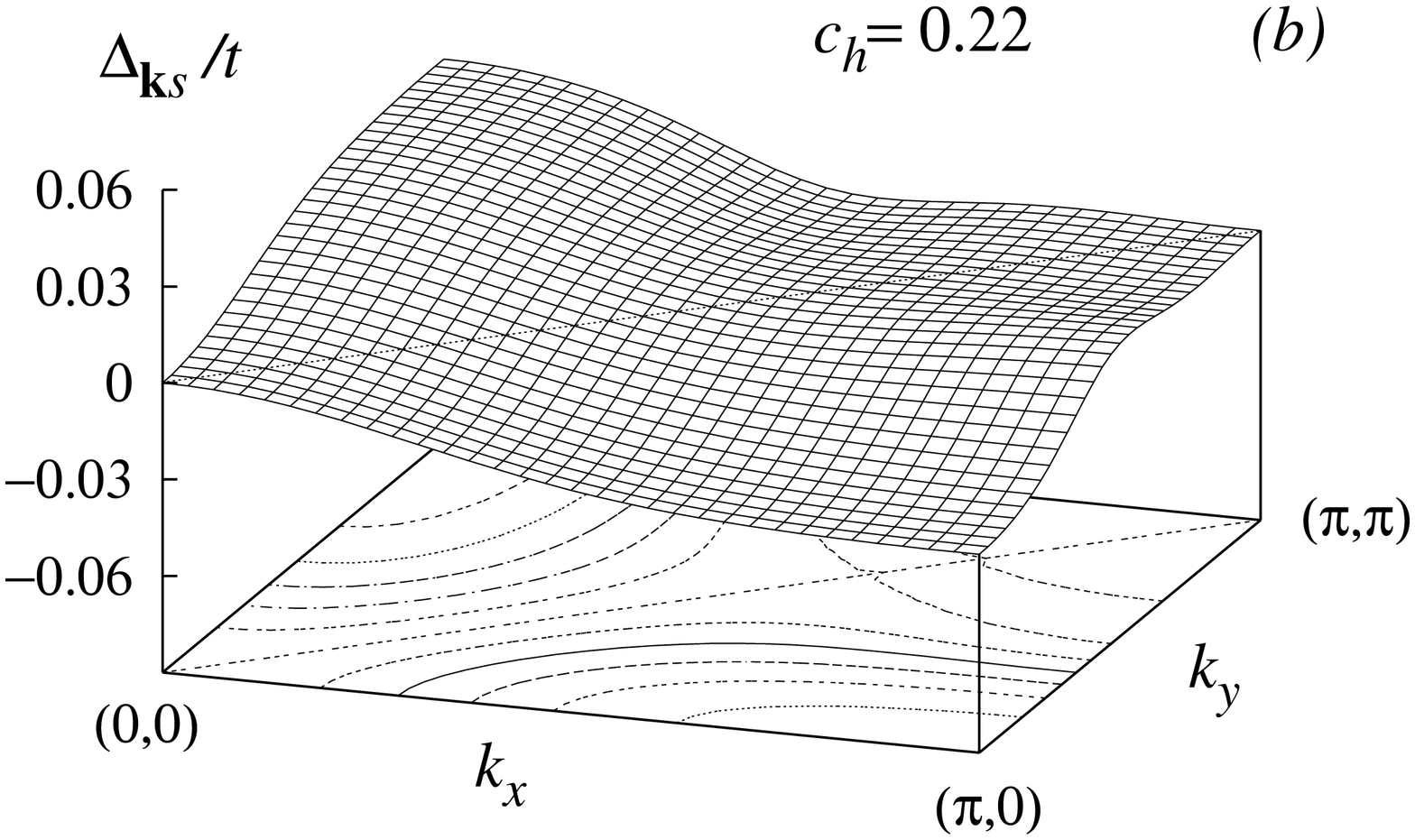}
   \end{tabular}
   \end{center}
   \caption[example] 
   { \label{fig:fig3} 
(a) Zero temperature $\Delta_{{\bf k}s}$ obtained from Eq.~(\ref{del}) using coupling
without MF term [as in Fig.~2, line (b)], for $c_h=0.12$ and  $t'/t=-0.3$. (b) As 
in (a), but for $c_h=0.22$.
}   \end{figure} 

As discussed above, we take here the dynamical spin susceptibility
independent of temperature, the approximation justified for low
temperature $T\sim T_c$ and solutions of the gap equation in the
$T<T_c$ regime should be considered only qualitatively.  In Fig.~4
full temperature dependence of $\Delta_{{\bf k}s}(T)$ is presented for
${\bf k}=(\pi,0)$ for various $c_h$ and $t'/t=-0.3$. An interesting
observation is increasing ratio $\Delta_{(\pi,0)s}(T=0)/T_c$ from OD to UD
regime reflecting different relative relevance of $J$- and $t'$-terms
in the coupling.

Spin-fermion coupling $\tilde m_{\bf kQ}$ is strongly
$t'$-dependent and as expected also $T_c$ exhibits a pronounced dependence on $t'/t$. 
In Fig.~5 we present results, as obtained for fixed OP
$c_h=0.17$, but different $t'/t<0$, as relevant for hole-doped
cuprates \cite{pava}.

   \begin{figure}
   \begin{center}
   \begin{tabular}{c}
   \includegraphics[height=7cm]{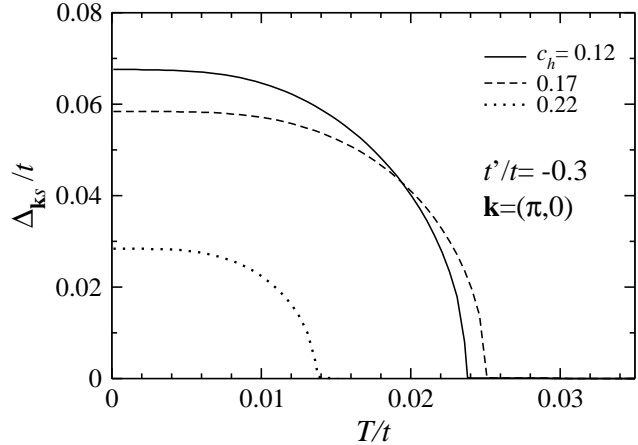}
   \end{tabular}
   \end{center}
   \caption[example] 
   { \label{fig:fig4} 
Temperature dependence of $\Delta_{{\bf k}s}(T)$ for ${\bf k}=(\pi,0)$ with
parameters and coupling corresponding to Fig.~3.
}
   \end{figure} 

   \begin{figure}
   \begin{center}
   \begin{tabular}{c}
   \includegraphics[height=7cm]{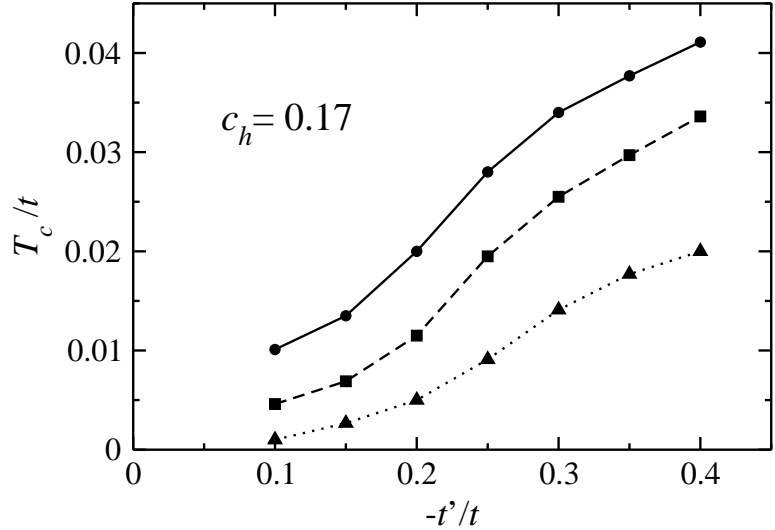}
   \end{tabular}
   \end{center}
   \caption[example] 
   { \label{fig:fig5} 
$T_c/t$ vs. $-t'/t$ for fixed 'optimum' doping $c_h=0.17$ and
different versions of Eq.~(\ref{del}), as in Fig.~2.
}
   \end{figure} 


\section{DISCUSSION}

Let us first comment on the relevance of the present method
and results to cuprates. Our starting point is the model,
Eq.~(\ref{tj}), where strong correlations are explicitly taken into
account via the projected fermionic operators. In this respect the
derivation crucially differs from the analysis of the phenomenological
spin-fermion model \cite{mont}. Nevertheless, in the latter approach
the resulting gap equation, Eq.~(\ref{del}), looks similar but
involves a constant effective coupling. In contrast, our $\tilde
m_{\bf kq}$, Eq.~(\ref{mkq}), is evidently ${\bf k},{\bf q}$-
dependent. In particular, in the most relevant region, i.e., along the
AFM zone boundary, $\tilde m_{\bf kQ}$ depends only on $t^\prime$ and
$J$, but not on $t$. This explains our central result novel within the
spin-fluctuation scenario, i.e., a pronounced dependence of $T_c$ on
$t^\prime$ which emerges directly via $t^\prime$ in the
effective interaction in Eq.~(\ref{tc}), and is consistent with the
evidence from different families of cuprates \cite{pava}. Similar
trend is obtained within the same model by the variational approach
\cite{shih}. One can give a plausible explanation of this effect. In
contrast to NN hopping $t$, the NNN $t^\prime$ represents the hopping
within the same AFM sublattice, consequently in a double unit cell
fermions couple directly to low-frequency AFM paramagnons, analogous
to the case of FM fluctuations generating superfluidity in $^3$He
\cite{legg}.

It is instructive to find an approximate BCS-like
formula which simulates our results. The latter involves the
characteristic cut-off energy $\Gamma_{\bf Q}$, while other relevant
quantities are the electron density of states ${\cal N}_0$ and $Z_m$
being the minimum $Z_{\bf k}$ on the FS (at the antinodal
point). Then, we get a reasonable fit to our numerical results with
the expression,
\begin{equation}
T_c \approx{1\over 2} \Gamma_{\bf Q} ~{\rm e}^{-2/( {\cal N}_0 V_{eff}) },
\label{tc}
\end{equation}
where the effective interaction is given by $V_{eff}= 3 Z_m
(2J-4t^\prime)^2 \chi_{\bf Q}$. Our numerical analysis suggests that
the main $t^\prime$-dependence of $T_c$ originates in the coupling
$\tilde m_{\bf kq}$, not in ${\cal N}_0 Z_m$, while the main
$c_h$-dependence comes from $\chi_{\bf Q}$ and $\Gamma_{\bf Q}$. Then,
Eq.~(\ref{tc}) implies that optimum doping, where $T_c$ reaches maximum,
increases with $-t'/t$. For parameters used in Fig.~1, e.g.,
$c_{\rm opt}=0.13+0.12(-t'/t)$.

In this analysis we do not extend our input data outside the doping
range $0.12<c_h<0.22$. Nevertheless, we can discuss on the basis of
Eq.~(\ref{tc}) the variation $T_c(c_h)$ elsewhere. Towards the undoped
AFM also the spin fluctuation scale should vanish $\Gamma_{\bf Q} \to
0$ and consequently $T_c(c_h \to 0) \to 0$. On the OD side, $\chi_{\bf
Q}$ and therefore $V_{eff}$ should decrease with doping, leading again to
fast reduction of $T_c(c_h)$.

It is evident from our analysis, that actual values of $T_c$ are quite
sensitive to input parameters and NS properties. Since we employ the
lowest-order mode-coupling approximation in a regime without a small
parameter, one can expect only a qualitatively correct behavior.
Still, calculated $T_c$ are in a reasonable range of values in
cuprates. We also note that rather modest 'optimum' $T_c$ value within
presented spin-fluctuation scenario emerge due to two competing
effects in Eqs.~(\ref{del}),(\ref{tc}): large $\tilde m_{\bf kq}$ and
$\chi_{\bf Q}$ enhance pairing, while at the same time through a
reduced $Z_{\bf k}$ and cutoff $\Gamma_{\bf Q}$ they limit $T_c$.

It should also be noted that in the UD regime we are dealing with the
strong coupling SC. Namely, we observe that ${\cal N}_0 V_{eff}$ shows
a pronounced increase at low doping mainly due to large $\chi_{\bf
Q}$.  Then it follows from Eq.~(\ref{tc}) that $T_c$ is limited and
determined by $\Gamma_{\bf Q}$. At the same time, INS experiments
\cite{fong} reveal that in the UD cuprates the resonant peak at $\omega
\sim \omega_r$ takes the dominant part of intensity of ${\bf q} \sim
{\bf Q}$ mode which becomes underdamped possibly even for
$T>T_c$. Thus it is tempting to relate $\Gamma_{\bf Q}$ to $\omega_r$
(for more extensive discussion see Ref.~\citenum{sega}) and in the UD
regime to claim $T_c \sim C \omega_r$, indeed observed in cuprates
\cite{fong} with $C \sim 0.26$.  However, additional work is needed to
accommodate properly an underdamped mode in our analysis.

Finally, let us conclude with a simplified qualitative picture emerging from
our numerical analysis.  As discussed above, full
numerical treatment reproduces doping- and $t'$-dependence consistent
with cuprates. NNN hoping matrix element $t'$ enters the formalism in
two ways. Firstly, through the dispersion of quasiparticles and
corresponding renormalisation in NS leading to moderate increase of
the density of states.  Contrary to some other previous works, this
effect is, surprisingly, not strongly doping dependent. More important
is the effect of the spin-fermion coupling, where $t'$ couples to spin
susceptibility. The main $t'$-dependence of $T_c$ comes from this
part. The doping dependence emerges mainly due to strong spin susceptibility
variation with $c_h$. Taking temperature independent
$\Gamma_{\bf Q}$
and $\chi_{\bf Q} \propto \Gamma_{\bf Q}^{-1}$ together with the assumption of
independence of the density of states on $c_h$, a simple
qualitative formula for $T_c$ emerges,

\begin{equation}
T_c = a \Gamma ~{\rm e}^{-b \Gamma/(c-t')^2 }.
\label{tcfinal}
\end{equation}

As an example we take $a\Gamma = (c_h-c_0)t$, $b=2a t$ and we neglect
spin exchange in the coupling, $c=0$.  In Fig.~5 doping dependence of
such $T_c$ is presented for various $t'/t$. In spite of severe
approximations used in the derivation of Eq.~(\ref{tcfinal}), this
result reflects the main features found also in the full numerical
treatment: i) $T_c$ is maximum in OP and decreasing in UD,OD regime,
ii) larger maximum $T_c$ at larger $-t'$, and iii) optimum $c_{\rm
opt}$ increases with $-t'$. In the UD regime $\Gamma$ is small
compared to the relevant temperature scale  and therefore for $c_h \sim c_0$
the use of a more precise form of $\Gamma(c_h,T)$ would be necessary. 
Maximum possible $T_c$
based on Eq.~(\ref{tcfinal}) is given with
$T_{\rm max}= 0.37 a\Gamma$, where $b\Gamma=(c-t')^2$
and therefore $c_{\rm opt}=c_0+a(c-t')^2/(bt)$.

Such estimates are consistent with our numerical results and could be
relevant also for real cuprate materials. Maximum possible transition
temperature is then given with approximate formula

\begin{equation}
T_{\rm max} \approx
{3\over 4{\rm e}}~ {\cal N}_0 Z_m (2J-4t^\prime)^2\chi_{\bf Q}\Gamma_{\bf Q}. 
\label{tcmax}
\end{equation}

   \begin{figure}
   \begin{center}
   \begin{tabular}{c}
   \includegraphics[height=7cm]{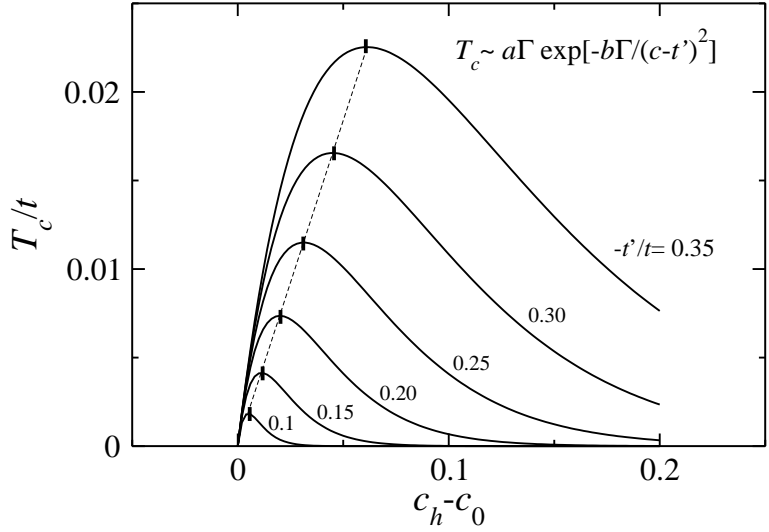}
   \end{tabular}
   \end{center}
   \caption[example] 
   { \label{fig:fig6} 
$T_c$ vs. $c_h$ with $a\Gamma = (c_h-c_0)t$, $b=2 a t$ and $c=0$. Dashed line
connects the positions of $T_{\rm max}=(c_{\rm opt}-c_0)t/{\rm e}$.
}
   \end{figure} 

\section{ACKNOWLEDGEMENTS}
Authors acknowledge stimulating discussions with N.M. Plakida and I. Sega.
This research was supported by the Ministry of Higher Education and Science of
Slovenia under grant P1-0044.





\end{document}